\documentclass[final,3p,times,twocolumn]{elsarticle}

\usepackage[utf8]{inputenc}
\usepackage[T1]{fontenc}
\usepackage{microtype} 
\usepackage[english]{babel}
\usepackage[utf8]{inputenx} 
\usepackage{hyperref}
\usepackage{appendix}
\usepackage{xcolor}
\usepackage{amsmath}
\usepackage{amssymb}
\usepackage{varioref}
\usepackage{wasysym}
\usepackage{verbatim}
\usepackage{lipsum}
\usepackage{xcolor, tikz, pgfplots}

\usetikzlibrary{matrix,calc,positioning,decorations.markings,decorations.pathmorphing,decorations.pathreplacing
}
\usetikzlibrary{arrows,cd,shapes}

\begin{document}

\begin{frontmatter}


\title{2-simplexes and superconformal central charges}

\author[label1,label2]{Federico Manzoni\fnref{1}}

\address[label1]{Mathematics and Physics department, Roma Tre, Via della Vasca Navale 84, Rome, Italy}
\address[label2]{INFN Roma Tre Section, Physics department, Via della vasca navale 84 Rome, Italy}

\fntext[1]{federico.manzoni@uniroma3.it}

\begin{abstract}
The superconformal central charge is an important quantity for theories emerging from geometrical engineering of Quantum Field Theory since it is linked, for example, to the scaling dimension of fields. Butti and Zaffaroni construction of the central charge for toric Calabi-Yau threefold geometries is a powerful tool but its implementation could be quite tricky. Here we present an equivalent new construction based on a 2-simplexes decomposition of the toric diagram.
\\
\\
\\
\textit{Published in Physics Letters B.}
\end{abstract}


\end{frontmatter}

\section{Introduction}
In the context of String Theory, the AdS/CFT correspondence represents a profound connection between the isometries of spacetime and the symmetry of a quantum field theory. In particular, the original formulation states that the local geometry around a stack of $N$ D3 branes in flat 10 dimensional spacetime has the form AdS$_5 \times S^5$ and the physics at large $N$ is holographically dual to the $\mathcal{N} =4$ 4d SCFT, defined on the boundary of the AdS spacetime \cite{Maldacena:1997re},\cite{ Gubser:1998bc},\cite{Witten:1998qj}. The geometry of the CY cone transverse to the branes, and in particular of its base referred to as horizon, determines the properties of the dual gauge side. The correspondence can be generalised to non-spherical horizons \cite{Morrison:1998cs},\cite{Klebanov:1998hh} $H^5$, a compact 5-dimensional Sasaki-Einstein space, so that the geometry is singular at the tip of the cone and some directions of the branes may be wrapped in compact cycles around the singularity\footnote{This context is known as geometrical engineering of Quantum Field theories and many dualities are been discovered using this tool, see ad example \cite{Antinucci:2020yki},\cite{Antinucci:2021edv},\cite{Manzoni:2022htx},\cite{Amariti:2022dui}.}. The local physics around the singularity results in a dual gauge side either $\mathcal{N}=2$ or $\mathcal{N}=1$. The $R$-symmetry of the SCFT plays a crucial role as its 't Hooft anomalies can be combined to give the central charge $a$
\begin{align}\label{eq:CentralCharge}
    a = \frac{3}{32} \left( 3 \mathrm{Tr }R^3 - \mathrm{Tr }R \right) \; ,
\end{align}
which represents a counting of the degrees of freedom of the field theory. The connection between global symmetry and geometry is expressed by the Gubser formula~\cite{Gubser_1998amax}
\begin{align}
    \mathrm{Vol}( H^5 ) = \frac{\pi^3}{4} \frac{N^2}{a} \; ,
\end{align}
where $N$ is the unit of 5-form flux supported by the compact horizon and $a$ is the central charge of the dual superconformal gauge theory. The $R$-charges of a general supersymmetric theory can not be uniquely assigned because any global symmetry factor can mix with the $U(1)_R$. However, in~\cite{Intriligator:2003jj} it is shown that the $R$-charges at the superconformal fixed point are unambiguously determined as those maximize central charge. 

The importance of the $R$-charges relies on the fact that the scaling dimension $\Delta$ of gauge-invariant operators depends on its $R$-charge $R$ as $\Delta = 3R/2$. The point is that we need to specify the geometry of the horizon in order to say what the associated SCFT is. A systematic construction of a singular CY cone has been developed in the case of a toric space, which is defined as a $\mathbb{T}^3$ fibration over a convex polygon. This type of cone has at least isometry $U(1)^3$ and the geometry is completely specified by the toric polygon or toric diagram. Given a toric diagram, there is a precise algorithm which allows one to construct the associated brane tiling \cite{Franco:2005sm},\cite{Franco_2006},\cite{ hanany2005dimer},\cite{Franco:2012mm},\cite{Franco:2013ana},\cite{2007iso} a bipartite graph that contains all the information about the gauge theory, i.e. gauge factors, matter fields and their interaction). This construction is powerful since from toric diagram and combinatorics we can obtain a field theory and we can analyze it with well established brane tiling techniques. A prominent example is the computation of the central charge using $a$-maximization \cite{Intriligator:2003jj}.

As the connection between isometry and symmetry works in both ways, one may expect that the computation of quantities in the gauge theory has a counterpart in the geometric side, and indeed $a$-maximization is equivalent to volume minimization \cite{Martelli:2005tp},\cite{Martelli:2006yb}. Moreover, in~\cite{Butti:2005vn} Butti and Zaffaroni developed an algorithm that allows us to compute the unique superconformal $R$-charges and the central charge of the SCFT directly from toric data. 
In this letter, after having reviewed briefly the algorithm given by Butti and Zaffaroni, we work out, in Section \ref{3.1}, a new procedure in terms of 2-simplexes decomposition of the toric diagram. This new procedure is in agreement with the Butti and Zaffaroni one, indeed each can be obtained starting from the other as shown in Section \ref{3.2}. In Section \ref{3.3} we link 2-simplexes to the Reeb vector field and we give an interpretation in terms of isoradial embedding.
In the end, we are going to give a working example and to discuss possible applications for future research.

\section{Central charge of toric gauge theories}\label{sec:ToricCharge}
In this section, we review the procedure to build up the superconformal $R$-charges and the central charge $a$ of a superconformal gauge theory due to Butti and Zaffaroni. First of all, let us discuss what information we need in the field theory side to build up the central charge $a$; then we are going to proceed with geometric side construction. 

Let us consider a gauge theory with gauge group $G=\prod_{i=1}^n \, SU(N)_i$ and a set $F$ of bifundamental matter superfields $X_{ij}$ that transform under the fundamental representation of $SU(N)_i$ and under the antifundamental representation of $SU(N)_j$. Each field $X_{ij}$ carries a charge $R_{ij}$ under the $R$-symmetry and the central charge $a$ is a combination of 't Hooft anomalies $\mathrm{Tr}\,R$ and $\mathrm{Tr}\,R^3$, see \eqref{eq:CentralCharge}. We are interested in SCFT holographically dual to a theory of gravity on AdS spacetime and holography requires that $\mathrm{Tr}\,R=0$ at large $N$; the contribution of all fermions yields 
\begin{align}
    a = \frac{9}{32} \left[ |G| + \sum_{X_{ij} \in F}\left( R_{ij} - 1  \right)^3 \right] \; ,
\end{align}
where the first term inside the parenthesis comes from gauginos contribution to the anomaly. A toric theory has at least $U(1)^2 \times U(1)_R$ global symmetry, where these abelian factors mix together. Thus $R_{ij}$, which are subject to the condition for anomaly cancellation, are a linear combination of charges, made of three variables. There may be other global abelian factors, whose mixed anomaly with $R$-symmetry vanish, i.e. $\mathrm{Tr}\,U(1)U(1)_R^2 = 0$; this type of global factors are called baryonic and do not mix with the $R$-symmetry~\cite{Intriligator:2003jj}. 

In \cite{Butti:2005vn} Butti and Zaffaroni showed how $a$-maximization can be performed by considering a point $\Vec{B}=(x,y)$ inside the polygon representing the toric diagram. In other words, the superconformal $R$-charges of a gauge theory associated to a toric geometry are determined by the point $\Vec{B}$ and toric data. The procedure is given in the following. \\
First, we define the product between two 2-dimensional vectors as 
\begin{align}
\langle u,v \rangle:=det \begin{bmatrix}
u^{(1)} & u^{(2)} \\
v^{(1)} & v^{(2)}
\end{bmatrix}
\; . \label{eq:uvdet}
\end{align}
For each extremal point in the toric diagram, we associate a vector $v_i$ going from vertex $i$ to vertex $i+1$, with $i=1,\,\ldots,\,d \ mod(d)$ where $d$ is the number of extremal points. The vectors $w_i$ are orthogonal to the $v_i$ and they define the $(p,q)$-web diagram, so the product $\langle {v_i,v_j}\rangle$ gives the entries in the adjacency matrix. For example, if $\langle{v_i,v_j}\rangle =2$ there are two fields connecting associated nodes in the quiver. \\
The next step is to define a set $C$, made by all positive $\langle{v_i,v_j}\rangle$. These are given by ordered pairs of vectors $(v_i, \, v_j )$ such that the associated $(p,q)$-web diagram vector $w_i$ is rotated counterclockwise to $w_j$ by an angle smaller than $\pi$. \\
At this point, to each vertex we associate a trial $R$-charge $a_i$ and to each element $(v_i, \, v_j )$ in the set $C$ we associate the trial $R$-charges combination $a_{i+1}+\, \ldots \,+ a_{j}$. This has a pictorial interpretation at the toric diagram level: moving a $(p,q)$-web vector $w_i$ to $w_j$, vertices from $i+1$ to $j$ are enclosed and so one picks up their trial charge. For example, if $\langle {v_1,v_3}\rangle =2$ then to the two fields a trial $R$-charge $a_2+a_3$ is given. As we know, the trial charges must satisfy the condition 
\begin{align}
\sum_{i=1}^d a_i = 2 \; . \label{eq:a-condition}
\end{align}
The final step is to build up the quantity
\begin{align}
a = \frac{9}{32} \left[ A_P + \sum_{(i,j)} |\langle {v_i,v_j}\rangle| \left( a_{i+1} + \ldots + a_j - 1 \right)^3 \right] \; , \label{eq:acharge}
\end{align}
where $A_P$ is the area of the polygon which represents the toric diagram; $(i,j) \in C$ and we must maximize it over the independent charges $a_i$. Here we can notice that it is strange that the polygon area appears while the $R$-charges are expressed not in terms of areas somehow.

This procedure gives us a way to count fields from toric data and associate them a trial $R$-charge; in principle this is a maximization that can involve more than two variables. However, Butti and Zaffaroni give an ansatz for each $a_i$ so that maximization of \eqref{eq:acharge} corresponds to minimization of the volume of the Sasaki-Einstein associated to the given toric geometry, hence reducing the number of variables down to two: the coordinates $(x,y)$ of a point $\Vec{B}$. We have to note that the point $\Vec{B}$ is nothing but the projection of the Reeb vector $\Vec{b}=3(1,x,y)$ over the $\mathbb{Z}^2$ lattice up to a numerical constant. 
Butti and Zaffaroni proposed that to each vertex of the polygon, we must associate another vector $r_i$ that connects a point $(x,y)$ inside the toric diagram to the vertex $i$. Then
\begin{align}
l_i(x,y) := \frac{\langle {v_{i-1},v_{i}} \rangle }{\langle {r_{i-1},v_{i-1}}\rangle \langle{r_{i},v_{i}}\rangle}, \label{eq:ldefinition}
\end{align}
and the trial charges are written as functions of $(x,y)$
\begin{align}
a_i(x,y) = 2 \frac{l_i(x,y)}{\sum_{k=1}^d l_k(x,y)} \label{eq:Rcharges} \;.
\end{align}
Inserting \eqref{eq:Rcharges} into \eqref{eq:acharge} we get the central charge $a$ in terms of the two coordinates of the Reeb vector on the toric diagram and the maximization of the central charge yields $(\Bar{x},\Bar{y})$, such that the theory is superconformal.

\begin{figure}[h]
\centering{
\begin{tikzpicture}[auto, scale=1.3]
        \node [circle, fill=black, inner sep=0pt, minimum size=2mm] (i-1) at (1,0) {};            
        \node [circle, fill=black, inner sep=0pt, minimum size=2mm] (i) at (0,1) {};
        \node [circle, fill=black, inner sep=0pt, minimum size=2mm] (q) at (0,0) {}; 
        \node [circle, fill=black, inner sep=0pt, minimum size=2mm] (i+1) at (0,-1) {}; 
        \node [right=0.2cm of i-1] {$i-1$};
        \node [left=0.2cm of i] {$i$};
        \node [left=0.2cm of q] {$q_i$};
        \node [left=0.2cm of i+1] {$i+1$};
         
        \draw (i-1) to node [swap] {$v_{i-1}$} (i) [->, thick, draw=black!20!red];
        \draw (i) to node [swap, pos=0.8] {$v_{i}$} (i+1) [->, thick, draw=black!60!blue];
\end{tikzpicture}}
\caption{An example of a side of a polygon with a non extremal point.}\label{fig:nonextremalexample}
\end{figure}
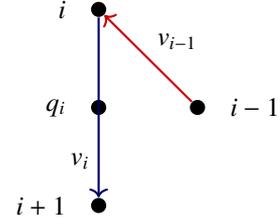

Let us now discuss what happens in case the polygon has some non extremal points. We denote such points as $q_i$ and we associate to them some trial $R$-charge $b_i$. Recall that $v_i$ are defined such that they connect two successive extremal points, that is there are some non extremal $q_i$, $v_i$ just pass over them to reach $v_{i+1}$; see Figure \ref{fig:nonextremalexample}. For a side with $q_{i}$, there are more than one vector $w$ of the $(p,q)$-web, but all of them are parallel. Suppose, as in Figure \ref{fig:nonextremalexample}, that a side of the polygon has vertices $i$ and $i+1$ with a $q$ in the middle; in moving a vector $w_{i-1}$ to $w_{i}$ we have two choices, namely stopping before or after the point $q$. These choices correspond to fields with trial charges $a_i$ and $a_i + b_i$, so we can assign to all field a trial charge with the condition
\begin{align}
\sum_{i=1}^d a_i + \sum_{j=1}^{\overline{d}} b_j = 2 \; , \label{eq:abcondition}
\end{align}
where $\overline{d}$ is the number of not extremal points. In \cite{Butti:2005vn}, it is pointed out that $a$-maximization sets all $b_i=0$, so Butti and Zaffaroni suggest that not extremal points are not relevant in determining the superconformal point of toric theories\footnote{This point has probably a more exhaustive and deeper geometrical meaning and justification which must be investigated in the future.}.

\section{2-simplexes decomposition and symplexic central charge}\label{2}
Let us start to discuss how to decompose the toric diagram into 2-simplexes and how to link the araes of these triangles to the trial $R$-charge, so to the central charge, thanks to equation \eqref{eq:acharge}. The basic idea is very simple: take the arbitrary point $\Vec{B}=(x,y)$ which corresponds to the Reeb vector projection on the toric diagram and connect this point with each vertex point of the diagram. In the previus section, we called these vectors $r_i$ and so we employed this notation also here; moreover, from Figure \ref{fig:ex}, it is obvious that the point $\Vec{B}=(x,y)$ must be inside the diagram. This construction for a generic toric diagram provides a set of areas $A_j$ given by the triangles delimited by $r_j$, $r_{j+1}$ and the toric diagram edge $v_j$, as reported in Figure \ref{fig:ex}.
\begin{figure}[h]
\centering{
\begin{tikzpicture}[auto, scale=2]
        \node [circle, fill=black, inner sep=0pt, minimum size=2mm] (p1) at (0,0) {};            
        \node [circle, fill=gray, inner sep=0pt, minimum size=2mm] (p2) at (2,1) {};
        \node [circle, fill=black, inner sep=0pt, minimum size=2mm] (p3) at (1,2) {}; 
        \node [circle, fill=gray, inner sep=0pt, minimum size=2mm] (s) at (1,1) {}; 
        \node [circle, fill=black!30!red, inner sep=0pt, minimum size=2mm] (b) at (1.3,0.9) {}; 
        \node [circle, fill=black, inner sep=0pt, minimum size=2mm] (a1) at (1,0) {}; 
        \node [circle, fill=black, inner sep=0pt, minimum size=2mm] (a2) at (2,0) {}; 
        \node [circle, fill=black, inner sep=0pt, minimum size=2mm] (a3) at (0,1) {};
        \node [circle, fill=gray, inner sep=0pt, minimum size=2mm] (a4) at (0,2) {};
        \node [circle, fill=gray, inner sep=0pt, minimum size=2mm] (a5) at (2,2) {};   
        \node [circle, fill=gray, inner sep=0pt, minimum size=2mm] (a6) at (3,0) {}; 
        \node [circle, fill=gray, inner sep=0pt, minimum size=2mm] (a8) at (0,3) {};
        \node [circle, fill=gray, inner sep=0pt, minimum size=2mm] (a9) at (0,3) {};
        \node [circle, fill=gray, inner sep=0pt, minimum size=2mm] (a10) at (3,3) {}; 
        \node [circle, fill=black, inner sep=0pt, minimum size=2mm]
        (a11) at (3,1) {}; 
        \node [circle, fill=black, inner sep=0pt, minimum size=2mm] (a12) at (3,2) {}; 
        \node [circle, fill=gray, inner sep=0pt, minimum size=2mm] (a16) at (1,3) {}; 
        \node [circle, fill=black, inner sep=0pt, minimum size=2mm] (a17) at (2,3) {};

        \draw (p1) to node [swap] {$v_1$} (a2) [->, thick];
        \draw (a2) to node [swap] {$v_2$} (a11) [->, thick];
        \draw (a11) to node [swap] {$v_3$} (a12) [->, thick];
        \draw (a12) to node [swap] {$v_4$} (a17) [->, thick];
        \draw (a17) to node [swap] {$v_5$} (a3) [->, thick];
        \draw (a3) to node [swap] {$v_6$} (p1) [->, thick];

        \draw (b) to node [swap, black!30!red] {$r_1$} (p1) [->, thick, draw=black!30!red];
        \draw (b) to node [swap, black!30!red] {$r_2$} (a2) [->, thick, draw=black!30!red];
        \draw (b) to node [black!30!red] {$r_3$} (a11) [->, thick, draw=black!30!red];
        \draw (b) to node [swap, black!30!red] {$r_4$} (a12) [->, thick, draw=black!30!red];
        \draw (b) to node [swap, black!30!red] {$r_5$} (a17) [->, thick, draw=black!30!red];
        \draw (b) to node [black!30!red] {$r_6$} (a3) [->, thick, draw=black!30!red];
        
        \node [above=0.35cm of a1] {$A_1$};
        \node [above=1.0cm of a2] {$A_2$};
        \node [above left=0.35cm of a11] {$A_3$};
        \node [left=1.2cm of a12] {$A_4$};
        \node [above=1.55cm of b] {$A_5$};
        \node [below right=0.45cm of a3] {$A_6$};
                
\end{tikzpicture}
\caption{Generic 2-simplexes decomposition of a generic toric diagram.}\label{fig:ex}
}
\end{figure}
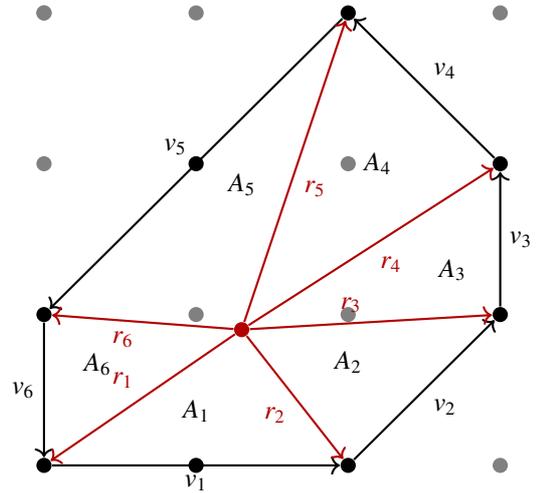\\
The area $A_j$ can be computed easily remembering that the toric diagram lies on a $\mathbb{Z}^2$ lattice plane and so vectors $v_j$ and $r_j$ do not have the third component $v_j^{(3)}=r_j^{(3)}=0$, hence
\begin{equation}
\begin{aligned}
A_j&=\frac{1}{2}\left| det
\begin{bmatrix} 
\hat{i} & \hat{j} & \hat{k} \\
r_j^{(1)} & r_j^{(2)} & 0  \\
v_j^{(1)} & v_j^{(2)} & 0
\end{bmatrix}
\right|=\frac{1}{2}\left|det \begin{bmatrix} 
r_j^{(1)} & r_j^{(2)}  \\
v_j^{(1)} & v_j^{(2)}
\end{bmatrix}
\right|.
\label{peppe32}
\end{aligned}
\end{equation}

\subsection{Rule for constructing the trial $R$-charges $a_i$ from areas $A_j$}\label{3.1}
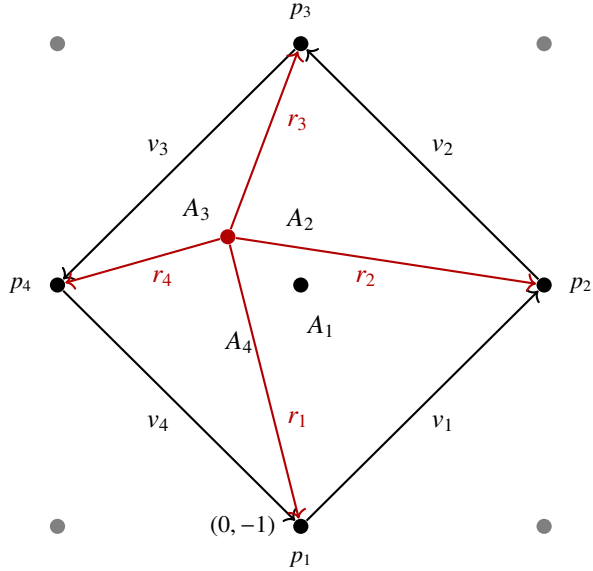
\begin{figure}[ht]
\centering{
\begin{tikzpicture}[auto, scale=3.2]
        \node [circle, fill=black, inner sep=0pt, minimum size=2mm] (p1) at (0,-1) {};            
        \node [circle, fill=black, inner sep=0pt, minimum size=2mm] (p2) at (1,0) {};
        \node [circle, fill=black, inner sep=0pt, minimum size=2mm] (p3) at (0,1) {}; 
        \node [circle, fill=black, inner sep=0pt, minimum size=2mm] (p4) at (-1,0) {};
        \node [circle, fill=black, inner sep=0pt, minimum size=2mm] (s) at (0,0) {}; 
        \node [circle, fill=black!30!red, inner sep=0pt, minimum size=2mm] (b) at (-0.3,0.2) {}; 

        \node [circle, fill=gray, inner sep=0pt, minimum size=2mm] (a2) at (-1,-1) {}; 
        \node [circle, fill=gray, inner sep=0pt, minimum size=2mm] (a3) at (1,-1) {};
        \node [circle, fill=gray, inner sep=0pt, minimum size=2mm] (a4) at (1,1) {};
        \node [circle, fill=gray, inner sep=0pt, minimum size=2mm] (a5) at (-1,1) {};   

        \draw (p1) to node [swap] {$v_1$} (p2) [->, thick];
        \draw (p2) to node [swap] {$v_2$} (p3) [->, thick];
        \draw (p3) to node [swap] {$v_3$} (p4) [->, thick];
        \draw (p4) to node [swap] {$v_4$} (p1) [->, thick];

        \draw (b) to node [swap, black!30!red, pos=0.7, swap] {$r_1$} (p1) [->, thick, draw=black!30!red];
        \draw (b) to node [swap, black!30!red] {$r_2$} (p2) [->, thick, draw=black!30!red];
        \draw (b) to node [black!30!red, swap, pos=0.7] {$r_3$} (p3) [->, thick, draw=black!30!red];
        \draw (b) to node [black!30!red] {$r_4$} (p4) [->, thick, draw=black!30!red];
                
        \node [below right=1.2cm of b] {$A_1$};
        \node [above=0.5cm of s] {$A_2$};
        \node [above left=0.03cm of b] {$A_3$};
        \node [below left=0.6cm of s] {$A_4$};
        \node [left=0.1cm of p1] {\small{$(0,-1)$}};
        \node [below=0.1 cm of p1] {\small{$p_1$}};
        \node [right=0.1 cm of p2] {\small{$p_2$}};
        \node [above=0.1 cm of p3] {\small{$p_3$}};
        \node [left=0.1 cm of p4] {\small{$p_4$}};        
                
\end{tikzpicture}
\caption{The toric diagram of $\mathbb{F}_0$.}\label{fig:F0Toric}
}
\end{figure}

Let us give the rule to build up all the trial $R$-charges $a_i$ in terms of the areas of the 2-simplexes $A_j$ of triangles constructed inside the toric diagram using the internal point $\Vec{B}=(x,y)$ as a vertex and $v_j$ as the base according to \eqref{peppe32}. Let us derive this rule with an example: the zeroth Hirzebruch surface ($\mathbb{F}_0$) whose toric diagram is in Figure \ref{fig:F0Toric}.
\\

From toric diagram above we can compute the vectors $v_i$ and $r_i$
\begin{align}
\begin{array}{lll}
v_1 = (1,1) \; , & \quad r_1 = (-x, -1-y) \; ; \\[5pt]
v_2 = (-1,1) \; , & \quad r_2 = (1-x, -y) \; ; \\[5pt]
v_3 = (-1,-1) \; , & \quad r_3 = (- x, 1-y) \; ; \\[5pt]
v_4 = (1,-1) \; , & \quad r_4 = (-1-x, -y) \; ; \\
\end{array} 
\end{align}
and from these we get the areas of the four triangles
\begin{equation}
\begin{aligned}
&A_1 = \frac{1}{2} \left( 1 - x + y \right);\\
&A_2 = \frac{1}{2} \left( 1 - x - y \right);\\
&A_3 = \frac{1}{2} \left( 1 + x - y \right);\\
&A_4 = \frac{1}{2} \left( 1 + x + y \right);
\end{aligned}    
\end{equation}
and the set $C$
\begin{align}
&\langle v_1,v_2 \rangle = 2  \; ; \nonumber \\[5pt]
&\langle v_4,v_1 \rangle = 2  \; ; \nonumber \\[5pt]
&\langle v_2,v_3 \rangle = 2  \; ; \nonumber \\[5pt]
&\langle v_3,v_4 \rangle = 2  \; .
\label{setC}
\end{align}
According to Butti-Zaffaroni procedure $R$-charges can be computed starting from \eqref{eq:Rcharges} using \eqref{eq:ldefinition}:
\begin{equation}
\begin{aligned}
a_1 &=  \frac{1}{2} \bigg( -x^2 + \left(y-1\right)^2 \bigg)= \\
&= \frac{16A_2 A_3}{8A_1A_2+8A_2A_3+8A_3A_4+8A_4A_1}\; ; \\
a_2 & = \frac{1}{2} \bigg( 1 + 2x + x^2 - y^2 \bigg)= \\
&= \frac{16A_3 A_4}{8A_1A_2+8A_2A_3+8A_3A_4+8A_4A_1}\; ; \\ 
a_3 & = \frac{1}{2} \bigg( -x^2 + (y+1) \bigg)= \\
&= \frac{16A_4 A_1}{8A_1A_2+8A_2A_3+8A_3A_4+8A_4A_1} \; ; \\
a_4 & = \frac{1}{2} \bigg( 1-2x+x^2-y^2  \bigg)= \\ 
&\frac{16A_1 A_2}{8A_1A_2+8A_2A_3+8A_3A_4+8A_4A_1} \; .
\end{aligned}    
\end{equation}
It can be noted that all expressions above are of the form
\begin{align}
    a_i = \frac{\langle {v_{i-1},v_i}\rangle}{D} \prod_{q\neq i,\, i-1} (2A_j) \; ,
\label{aaree}    
\end{align}
where the product involves all areas $A_j$ which do not have $v_i$ nor $v_{i-1}$ as an edge\footnote{This is the meaning of $q \neq i,i-1$.} and $D$ is a combination of all the areas $A_j$; in this case $D=8A_1A_2+8A_2A_3+8A_3A_4+8A_4A_1$. This explicit expression for $D$ can be recovered in fully generality using ansatz \eqref{aaree} and the constraint $\sum_{k=1}^d a_k=2$ where $d$ is the number of extremal points of the diagram:
\begin{equation}
    \sum_{k=1}^d \frac{1}{D}\langle {v_{k-1},v_k}\rangle \prod_{q\neq k,\, k-1} (2A_q) = 2 \; ,
\end{equation}   
which gives
\begin{equation}
    D=\frac{1}{2}\sum_{k=1}^d\langle {v_{k-1},v_k}\rangle \prod_{q\neq k,\, k-1} (2A_q) \; ,
\label{denominatorD}    
\end{equation}
since $D$ does not depend on $k$ because it is a combination of all the areas; this is claimed looking at several examples. From \eqref{denominatorD} and using relations \eqref{setC}, we get exactly the right denominator $D=8A_1A_2+8A_2A_3+8A_3A_4+8A_4A_1$.
Plugging \eqref{denominatorD} back in \eqref{aaree} we get the general form of the trial $R$-charges
\begin{equation}
    a_i=2\frac{\langle {v_{i-1},v_i}\rangle \prod_{q\neq i,\, i-1} (2A_q)}{\sum_{k=1}^d\langle {v_{k-1},v_k}\rangle \prod_{q\neq k,\, k-1} (2A_q)};
\label{eq:RchargesArea}    
\end{equation}
these $R$-charges do the right job in all cases, as can be seen looking at the formal equivalence between this procedure and the one proposed by Butti and Zaffaroni, in Paragraph \ref{3.2}.

At this point, the trial $R$-charges are related to the areas of the 2-simplexes with which we have decomposed the toric diagram; moreover, the area of the toric diagram is nothing but the sum of all the triangles areas $A_P=\sum_{k=1}^d A_k$ and so the central charge is completely determined by the areas $A_j$. The central charge \eqref{eq:acharge} reads
\begin{equation}
\begin{aligned}
    a&=\frac{9}{32}\bigg[\sum_{k=1}^d A_k +\sum_{(i,j)} \langle v_i,v_j \rangle \times \\
    & \bigg(2 \frac{\sum_{s=i+1}^j\langle {v_{s-1},v_{s}}\rangle \prod_{q\neq s,\, s-1} (2A_q)}{\sum_{k=1}^d\langle {v_{k-1},v_k}\rangle \prod_{q\neq k,\, k-1} (2A_q)} - 1 \bigg)^3\bigg],
\label{ccta}
\end{aligned}
\end{equation}
since these areas are function of the point $(x,y)$ the central charge is a function of only these two variables, in agreement with the Butti and Zaffaroni construction.

We conclude this paragraph with a comment on the structure of the $a_i$: we can think to $Z=\sum_{k=1}^d\langle {v_{k-1},v_k}\rangle \prod_{q\neq k,\, k-1} (2A_q)$ as the point dependent partition function of the toric diagram while $\langle {v_{i-1},v_i}\rangle \prod_{q\neq i,\, i-1} (2A_q)$ is the single realization; hence, we can think the trial central charge $a_i(x,y)$ as some probability density apart from factor two which can be reabsorbed in a new definition $P_i(x,y)=\frac{a_i(x,y)}{2}$.

\subsection{From Butti-Zaffaroni to symplexic charges}\label{3.2}
Let us show the equivalence between Butti and Zaffaroni trial charges, equation \eqref{eq:Rcharges}, and the symplexic charges \eqref{eq:RchargesArea}. Starting from \eqref{eq:Rcharges}, using \eqref{eq:ldefinition} and some algebra, we get
\begin{equation}
\begin{aligned}
a_i&=2\frac{\frac{\langle v_{i-1},v_i \rangle}{\langle r_{i-1},v_{i-1} \rangle \langle r_{i},v_{i} \rangle}}{\sum_{k=1}^d \frac{\langle v_{k-1},v_k \rangle}{\langle r_{k-1},v_{k-1} \rangle \langle r_{k},v_{k} \rangle}}=\\
&=2\frac{\frac{\langle v_{i-1},v_i \rangle}{\langle r_{i-1},v_{i-1} \rangle \langle r_{i},v_{i} \rangle}}{ \frac{\sum_{k=1}^d \langle v_{k-1},v_k \rangle \prod_{q \neq k,k-1}\langle r_{k},v_{k} \rangle}{\prod_{q=1}^d\langle r_{k},v_{k} \rangle}}=\\
&=2\frac{\langle v_{i-1},v_i \rangle \prod_{q \neq i,i-1}\langle r_{q},v_{q} \rangle}{\sum_{k=1}^d \langle v_{k-1},v_k \rangle \prod_{q \neq k,k-1}\langle r_{q},v_{q} \rangle};
\end{aligned}    
\end{equation}
now using \eqref{peppe32} we arrive to equation \eqref{eq:RchargesArea}
\begin{equation}
    a_i=2\frac{\langle v_{i-1},v_i \rangle \prod_{q \neq i,i-1} (2A_q)}{\sum_{k=1}^d \langle v_{k-1},v_k \rangle \prod_{q \neq k,k-1} (2A_q)};
\end{equation}
this writing simplified a lot the determination of trial $R$-charges. The equivalence between Butti and Zaffaroni procedure and the symplexic decomposition procedure makes it manifest that all anomalies cancellation discussed in the appendices of \cite{Butti:2005vn} are valid in this approach too. 

\subsection{Reeb vector, areas,  $R$-charges and isoradial embedding} \label{3.3}
In this section we want to point out the link between the areas and the work integral of the projected Reeb vector $\Vec{B}$; in fact from the point $\Vec{B}=(x,y)$ we can write down an expression for the toric diagram area and for triangles' areas. \\
The area $A_P$ of the toric diagram $P$ can be written in term of the area 2-form $\omega=dx \wedge dy$:
\begin{equation}
    A_P=\int_{\Omega}\alpha,
\end{equation}
where $\Omega=\{(x,y) \in P\}$. From $\omega$ we can find a one-form $\alpha$ such that $d\omega=\alpha$: the result is $\alpha=\frac{xdy-ydx}{2}$; indeed $\omega=d\alpha=\frac{dx \wedge dy}{2}-\frac{dy \wedge dx}{2}=dx \wedge dy$. At this 1-form is associated a vector field with component $\frac{1}{2}(-y,x)$ and we note that this is the point $\Vec{B}$ after the transformation belonging to $SL(2,\mathbb{Z})$ given by the matrix\footnote{Toric diagrams are defined up to an $SL(2,\mathbb{Z})$ tranformation and so physics is $SL(2,\mathbb{Z})$ invariant.} 
\begin{equation}
M=\begin{bmatrix}
0 & -1 \\
1 &  0
\end{bmatrix}.   
\end{equation}
A crucial point to note is that this $M \in SL(2,\mathbb{Z})$ transformation does not modify neither the area of the polygon nor the areas of the 2-simplexes since it is a $\frac{\pi}{2}$-rotation, but in this toric phase the projected Reeb vector is exactly the vector field associated to the 1-form $\alpha$ up to a numerical constant.
Hence polygon's area is given by
\begin{equation}
    A_P=\int_{\Omega}\omega=\int_{\Omega}d\alpha \underbrace{=}_{Stokes' \ THM} \int_{\partial \Omega} \alpha,
\end{equation}
where $\partial \Omega$ is the boundary of the polygon and so its edges. Splitting the integral over $\partial \Omega$ in a sum of integral over each edge $S_i$, we get
\begin{equation}
\begin{aligned}
    A_P&=\sum_{i}\int_{S_i}\frac{1}{2}(xdy-ydx)=\\
    &=\frac{1}{2}\sum_i\int_{S_i}(-y,x) \cdot (dx,dy)=\\
    &=\frac{1}{2}\sum_i\int_{S_i}\Vec{B} \cdot \Vec{dl}:=\frac{1}{2}\int_{\Gamma}\Vec{B} \cdot \Vec{dl},
\label{eq10}  
\end{aligned}
\end{equation}
where $\Gamma$ is the curve that enclose the toric diagram. At the same manner, we can write the area of a single triangle, $A_j$, using the projected Reeb vector:
\begin{equation}
    A_{j}=\frac{1}{2}\sum_i\int_{s_i} \Vec{B} \cdot \Vec{dl}:=\frac{1}{2}\int_{\gamma_q} \Vec{B} \cdot \Vec{dl} ,
    \label{giacomino}
\end{equation}
where now $s_i$ are the edges of the triangle and $\gamma_q$ is the curve that enclose the are $A_j$. These relations are intuitive since if we move the point $\Vec{B}$ we modify the areas $A_j$ as well.\\
Moreover, we know that every $a_i(x,y)$ can be expressed in terms of the triangles' areas and so they are intimately related to the projected Reeb vector' work integral:
\begin{align}\label{aprvwi}
    a_i=2\frac{\langle {v_{i-1},v_i}\rangle \prod_{q\neq i,\, i-1} \int_{\gamma_q} \Vec{B} \cdot \Vec{dl}}{\sum_{k=1}^d[\langle {v_{k-1},v_k}\rangle \prod_{q\neq k,\, k-1}\int_{\gamma_q} \Vec{B} \cdot \Vec{dl}]}.
\end{align}
We can give an interpretation of these at the level of brane tiling \cite{Franco:2005sm},\cite{ Franco_2006},\cite{hanany2005dimer},\cite{ Franco:2012mm},\cite{Franco:2013ana} thanks to isoradial embedding \cite{2007iso} where every $R$-charge is thought as an angle. Let us call $R_j$ the $R$-charge of a field $j$ around a vertex of the brane tiling; since every $R$-charge of a field is a combination of some trial $R$-charge \eqref{aprvwi}, $R_j$ reads
\begin{equation}
    R_j=\sum_{l_j}a_{l_j}=\frac{\sum_{l_j}\langle {v_{l_j-1},v_{l_j}}\rangle \prod_{q\neq l_j,\, l_j-1} \int_{\gamma_q} \Vec{B} \cdot \Vec{dl}}{D},
\end{equation}
where we make use of \eqref{denominatorD} and \eqref{giacomino}; isoradial embedding angle is recovered multiplying by $\pi$ both members. We can imagine that every term in the RHS sum contributes with an angle and that their sum gives the isoradial embedding angle $\theta_j=\pi R_j$, hence:
\begin{equation}
    \sum_{l_j}\frac{\theta_j}{n_{l_j}}=\frac{\pi \sum_{l_j}\langle {v_{l_j-1},v_{l_j}}\rangle \prod_{q\neq l_j,\, l_j-1} \int_{\gamma_q} \Vec{B} \cdot \Vec{dl} }{D};
\end{equation}
with the consistency condition $\sum_{l_j}\frac{1}{n_{l_j}}=1$. By eliminating sums, rearranging terms and use Fubini theorem\footnote{Using an explicit parametrization of the $q$-th curve in term of a parameter $t_q$, we get a product of ordinary integrals that can be rewritten as a $q$-dimensional integral on the domain, given by the product of the single domains of parameters $t_q$, and with integrand given by the product of the single integrand.} we get
\begin{equation}
    n_{l_j}=\frac{D \theta_j}{\langle {v_{l_j-1},v_{l_j}}\rangle \pi}\frac{1}{ \int_{V} \prod_{q\neq l_j,\, l_j-1} [\Vec{B}(t_q) \cdot \gamma(t_q)dt_q]},
    \label{part}
\end{equation}
where $V$ is the $q$-dimensional integration domain. The isoradial embedding angle $\theta_j$ is partitioned into smaller angles; how much the single $a_{l_j}$ counts in the partitioning is expressed by \eqref{part}: the larger are the areas enter in $a_{l_j}$, the smaller is $n_{lj}$ and the greater is the contribution to the isoradial embedding angle $\theta_j$, as we expected.

\section{Examples}\label{4}
Let us report two examples as working gym to build up symplexic central charge.
\subsection{Example 1: $\frac{\mathbb{C}^3}{\mathbb{Z}_3}$}
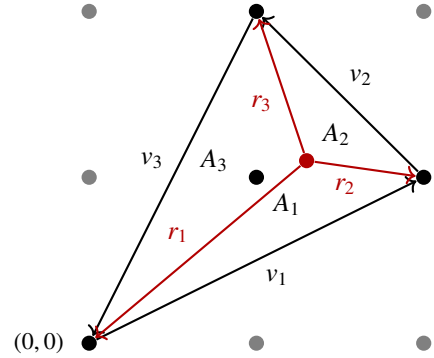
\begin{figure}[ht]
\centering{
\begin{tikzpicture}[auto, scale=2.2]
        \node [circle, fill=black, inner sep=0pt, minimum size=2mm] (p1) at (0,0) {};            
        \node [circle, fill=black, inner sep=0pt, minimum size=2mm] (p2) at (2,1) {};
        \node [circle, fill=black, inner sep=0pt, minimum size=2mm] (p3) at (1,2) {}; 
        \node [circle, fill=black, inner sep=0pt, minimum size=2mm] (s) at (1,1) {}; 
        \node [circle, fill=black!30!red, inner sep=0pt, minimum size=2mm] (b) at (1.3,1.1) {}; 
        \node [circle, fill=gray, inner sep=0pt, minimum size=2mm] (a1) at (1,0) {}; 
        \node [circle, fill=gray, inner sep=0pt, minimum size=2mm] (a2) at (2,0) {}; 
        \node [circle, fill=gray, inner sep=0pt, minimum size=2mm] (a3) at (0,1) {};
        \node [circle, fill=gray, inner sep=0pt, minimum size=2mm] (a4) at (0,2) {};
        \node [circle, fill=gray, inner sep=0pt, minimum size=2mm] (a5) at (2,2) {};   

        \draw (p1) to node [swap] {$v_1$} (p2) [->, thick];
        \draw (p2) to node [swap] {$v_2$} (p3) [->, thick];
        \draw (p3) to node [swap] {$v_3$} (p1) [->, thick];

        \draw (b) to node [swap, black!30!red] {$r_1$} (p1) [->, thick, draw=black!30!red];
        \draw (b) to node [swap, black!30!red] {$r_2$} (p2) [->, thick, draw=black!30!red];
        \draw (b) to node [black!30!red] {$r_3$} (p3) [->, thick, draw=black!30!red];
        
        \node [below right=0.03cm of s] {$A_1$};
        \node [above right=0.01cm of b] {$A_2$};
        \node [left=0.8cm of b] {$A_3$};
        \node [left=0.1cm of p1] {\small{$(0,0)$}};
                
\end{tikzpicture}
\caption{Toric diagram of $\frac{\mathbb{C}^3}{\mathbb{Z}_3}$.}\label{fig:C3Z3Toric}
}
\end{figure}

We start with a simple abelian orbifold of $\mathbb{C}^3$, its toric diagram is reported in Figure \ref{fig:C3Z3Toric}.
 \\
Quantities of interest are
\begin{align}
\begin{array}{lll}
v_1 = (2,1) \; , & \quad r_1 = (2-x, 1-y); \; \\[5pt]
v_2 = (-1,1) \; , & \quad r_2 = (1-x, 2-y); \; \\[5pt]
v_3 = (-1,-2) \; , & \quad r_3 = (-x, -y); \;
\end{array} 
\end{align}
and areas are
\begin{equation}
\begin{aligned}
&A_1 = \frac{1}{2} \left( 2y - x \right); \\
&A_2 = \frac{1}{2} \left( 3 - x - y \right); \\
&A_3 = \frac{1}{2} \left( 2x - y \right).
\end{aligned}
\end{equation}
The set $C$ contains
\begin{align}
&\langle{v_1,v_2}\rangle = 3  \; ; \nonumber \\[5pt]
&\langle{v_2,v_3}\rangle = 3  \; ; \nonumber \\[5pt]
&\langle{v_3,v_1}\rangle = 3  \; ; \nonumber 
\end{align} 
and using  equations \ref{eq:RchargesArea} and \ref{ccta} (or equivalently \ref{eq:acharge}) we obtain the trial $R$-charges
\begin{align}
a_1 & = \frac{2 A_2}{A_1+A_2+A_3} = \frac{2}{3} \left( 3 - x - y \right)\; ; \nonumber \\[5pt]
a_2 & = \frac{2 A_3}{A_1+A_2+A_3}= \frac{2}{3} \left( 2x - y \right) \; ; \nonumber \\[5pt]
a_3 &= \frac{2 A_1}{A_1+A_2+A_3} = \frac{2}{3} \left( 2y - x \right) \; .
\end{align}
and the central charge $a$,
\begin{equation}
\begin{aligned}
&a=  \frac{9}{32} \bigg[ \frac{3}{2} +3\left( \frac{2}{3} \left( 2x - y \right) - 1 \right)^3 +\\
&3\left( \frac{2}{3} \left( 2y - x \right) - 1 \right)^3 +3\left( \frac{2}{3} \left( 3 - x - y \right) - 1 \right)^3 \bigg]= \\
& \ \ = \frac{3}{64} [ 32 y^3 -9 - 48 y ( x-5) x - 96 x^2 +\\
& \ \ \ \ \ \ \ \ \ \ \ 32 x^3 - 48 y^2 (2 + x)].
\end{aligned}
\end{equation}
Note that this construction fixes the convention for the area of the elementary triangle $A_T$, those that represent the toric diagram for $\mathbb{C}^3$, while in the Butti and Zaffaroni construction this is not fixed\footnote{This means that we can choose the normalization $A_T=\frac{1}{2}$ or $A_T=1$. This has an impact on the value of the toric diagram area that appears in Butti and Zaffaroni central charge but this it does not alter the maximization procedure since it is only a constant}. The area is fixed to $A_T=\frac{1}{2}$. This function has a maximum at $(\Bar{x},\Bar{y})=(1,1)$ and all the $R$-charges are $a_i=\frac{2}{3}$. This is exactly the result we would get if we implemented the the central charge using Butti and Zaffaroni procedure. \\
\subsection{Example 2: SPP}
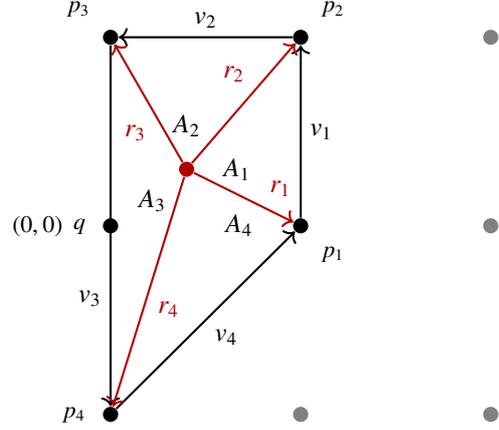
\begin{figure}[ht]
\centering{
\begin{tikzpicture}[auto, scale=2.5]
        \node [circle, fill=black, inner sep=0pt, minimum size=2mm] (p1) at (1,0) {};            
        \node [circle, fill=black, inner sep=0pt, minimum size=2mm] (p2) at (1,1) {};
        \node [circle, fill=black, inner sep=0pt, minimum size=2mm] (p3) at (0,1) {}; 
        \node [circle, fill=black, inner sep=0pt, minimum size=2mm] (q) at (0,0) {};
        \node [circle, fill=black, inner sep=0pt, minimum size=2mm] (p4) at (0,-1) {}; 
        \node [circle, fill=black!30!red, inner sep=0pt, minimum size=2mm] (b) at (0.4,0.3) {}; 

        \node [circle, fill=gray, inner sep=0pt, minimum size=2mm] (a2) at (1,-1) {}; 
        \node [circle, fill=gray, inner sep=0pt, minimum size=2mm] (a3) at (2,-1) {};
        \node [circle, fill=gray, inner sep=0pt, minimum size=2mm] (a4) at (2,0) {};
        \node [circle, fill=gray, inner sep=0pt, minimum size=2mm] (a5) at (2,1) {};   

        \draw (p1) to node [swap] {$v_1$} (p2) [->, thick];
        \draw (p2) to node [swap] {$v_2$} (p3) [->, thick];
        \draw (p3) to node [swap, pos=0.7] {$v_3$} (p4) [->, thick];
        \draw (p4) to node [swap] {$v_4$} (p1) [->, thick];

        \draw (b) to node [swap, black!30!red, pos=0.67, swap] {$r_1$} (p1) [->, thick, draw=black!30!red];
        \draw (b) to node [swap, black!30!red, swap, pos=0.6] {$r_2$} (p2) [->, thick, draw=black!30!red];
        \draw (b) to node [black!30!red, pos=0.4] {$r_3$} (p3) [->, thick, draw=black!30!red];
        \draw (b) to node [black!30!red] {$r_4$} (p4) [->, thick, draw=black!30!red];
                
        \node [left=0.4cm of p1] {$A_4$};
        \node [right=0.25cm of b] {$A_1$};
        \node [above=0.2cm of b] {$A_2$};
        \node [below left=0.1cm of b] {$A_3$};
        \node [left=0.4cm of q] {\small{$(0,0)$}};
        \node [left=0.1cm of q] {\small{$q$}};        
        \node [below right=0.1 cm of p1] {\small{$p_1$}};
        \node [above right=0.1 cm of p2] {\small{$p_2$}};
        \node [above left=0.1 cm of p3] {\small{$p_3$}};
        \node [left=0.1 cm of p4] {\small{$p_4$}};        
                
\end{tikzpicture}
\caption{Toric diagram of SPP.}\label{fig:SPPToric}
}
\end{figure}

Consider now the Suspended Pinch Point (SPP) singularity, the toric diagram is drown in figure \ref{fig:SPPToric}.
\\

The interesting quantities are
\begin{align}
\begin{array}{lll}
v_1 = (0,1) \; , & \quad r_1 = (1-x, -y) \; ; \\[5pt]
v_2 = (-1,0) \; , & \quad r_2 = (1-x, 1-y) \; ; \\[5pt]
v_3 = (0,-2) \; , & \quad r_3 = (- x, 1-y) \; ; \\[5pt]
v_4 = (1,1) \; , & \quad r_4 = (-x, -1-y) \; ; \\
\end{array} 
\end{align}
and the areas are 
\begin{equation}
\begin{aligned}
&A_1 = \frac{1}{2} \left( 1 - x \right); \\
&A_2 = \frac{1}{2} \left( 1 - y \right); \\
&A_3 = x; \\
&A_4 = \frac{1}{2} \left( 1 - x + y \right); 
\end{aligned}    
\end{equation}
The set $C$ is given by
\begin{align}
&\langle{v_1,v_2}\rangle = 1 \; ; \nonumber \\[5pt]
&\langle{v_2,v_3}\rangle = 2 \; ; \nonumber \\[5pt]
&\langle{v_4,v_2}\rangle = 1 \; ; \nonumber \\[5pt]
&\langle{v_3,v_4}\rangle = 2 \; ; \nonumber \\[5pt]
&\langle{v_4,v_1}\rangle = 1 \; . 
\end{align}
Trial $R$-charges are
\begin{equation}
\begin{aligned}
a_1 &= \frac{2A_2 A_3}{A_2 A_3 + A_3 A_4 + 2 A_4 A_1 + 2 A_1 A_2} = \\
&=\frac{2}{2-x} x \left( 1 - y \right)  \; ;  \\
a_2  &= \frac{2A_3 A_4}{A_2 A_3 + A_3 A_4 + 2 A_4 A_1 + 2 A_1 A_2} =\\ &=\frac{2}{2-x} x \left( 1 - x +y \right)\; ;  \\
a_3 &= \frac{4 A_4 A_1}{A_2 A_3 + A_3 A_4 + 2 A_4 A_1 + 2 A_1 A_2}= \\ &=\frac{2}{2-x} \left( 1 - x \right)\left( 1-x+y \right)   \; ; \\
a_4  &= \frac{4 A_1 A_2}{A_2 A_3 + A_3 A_4 + 2 A_4 A_1 + 2 A_1 A_2} =\\
&=\frac{2}{2-x} \left( 1 - x \right)\left( 1 - y \right)\; ;
\end{aligned}    
\end{equation}
the central charge $a$, from equation \ref{ccta} has a maximum at 
\begin{align}
& \Bar{x} = 1 - \frac{1}{\sqrt{3}} \; ; \nonumber \\[5pt]
& \Bar{y} = \frac{1}{2} \left( 1 - \frac{1}{\sqrt{3}} \right) \; ;
\end{align}
where
\begin{align}
    & a_1 = a_2 = 1 - \frac{1}{\sqrt{3}} \; ; \nonumber \\[5pt]
    & a_3 = a_4 = \frac{1}{\sqrt{3}} \; 
\end{align}

\section{Conclusions}
The central charge for theories arising from toric CY threefold can be computed from a completely combinatoric procedure since toric geometry is essentially combinatorics. Procedure presented here is based on a 2-simplexes decomposition of the toric diagram and the central charge is written only in term of areas of these simplexes using \eqref{ccta}. Areas of these triangles are easily computed thanks to relation \eqref{peppe32} and then trial $R$-charges are constructed according to \eqref{eq:RchargesArea}. On the one hand simplexes procedure makes the link between the triangles areas and the central charge more evident. On the other hand the time cost is reduced compared to the Butti and Zaffaroni procedure: for the generic example of Figure \ref{fig:ex}, computation time for a single trial $R$-charge is reduced from 0.008376$s$ to 0.001536$s$, about 18\% of the time needed to run out the Butti and Zaffaroni procedure.

An interesting point to be pointed out is that these areas can also be calculated using contour integral of the Reeb vector field and that they can be linked to the isoradial embedding angle, providing an interpretation at brane tiling level; it can be better understood and extended to the five brane system in a future work. 

Two last comments are in order. We see in Section \ref{3.1} that we can interpret $Z=\sum_{k=1}^d\langle {v_{k-1},v_k}\rangle \prod_{q\neq k,\, k-1} (2A_q)$ as the point dependent partition function of the toric diagram. Thermodynamics  variables associated to this partition function can contain information about the geometry of the toric CY variety and so of the field theory. The computation of the thermodynamic variables associated with this partition function could be the subject of a future work.\\
The final point is about a possible generalization to other models. The construction in term of 2-simplexes decomposition holds for (3+1)-dimensional theories but it looks easy to adapt to cases of different dimensions contrary to the procedure of Butti and Zaffaroni. The crucial difference lies in the dimension of the Sasaki-Einstein base, and so in the dimension of the CY cone that we need to get the 10-dimensional background space. Indeed if we want a (5+1)-dimensional superconformal field theory we have to embed D5-branes into a $\mathbb{M}^{5,1} \times \mathrm{CY}_4$ space-time: so this CY cone has 2 complex dimensions and its toric diagram is a segment; 2-simplexes are now 1-simplexes and areas are now segments. In geometrical engineering of QFTs with a $2n$-dimensional superconfarmal field theory, as the one ($n=2$) treated in this work, we need a CY cone of dimension $10-2n$ and so its toric diagram is a $(\frac{10-2n}{2}-1)$-dimensional object; 2-simplexes now are $(\frac{10-2n}{2}-1)$-simplexes. The possibility of extending the procedure to field theories with $d\neq 4$ can be approached in future works also studying whether such procedures can be applied to odd-dimensional field theories emerging from M-theory.

\subsubsection*{Acknowledgment}
This work was born from the Master's Degree thesis of Federico Manzoni collaborating with Salvo Mancani, PhD student of Fabio Riccioni. I thank Professor Fabio Riccioni, as master thesis advisor, and Professor Massimo Bianchi for interesting discussions.

\bibliographystyle{elsarticle-num} 
\bibliography{biblio}

\end{document}